# Probing three-dimensional structures of complex colloidal quantum dots at the single-atomic level

Qikai Wu[1†], Meng Pei[2†], Jiancheng Zhang[3], Wei Xu[2], Tianding Xu[1], Colin Ophus[4*], Zaiping Zeng[2*], Botao Ji[3,5*], Yao Yang[1*]

[1]*Department of Materials Science and Engineering, School of Engineering, Westlake University; Hangzhou, 310030, China.* [2]*Henan International Joint Laboratory of Quantum Dot Materials and School of Nanoscience and Materials Engineering, Henan University; Henan, 475001, China.* [3]*Zhejiang Key Laboratory of 3D Micro/Nano Fabrication and Characterization, Department of Electronic and Information Engineering, School of Engineering, Westlake University, Hangzhou, 310030, China.* [4]*Department of Materials Science and Engineering, Stanford University, Stanford, California 94305, United States* [5]*Westlake Institute for Optoelectronics, Hangzhou, 311421, China.*

[†]*These authors contributed equally to this work.*

**Colloidal quantum dots (QDs) are promising optoelectronic materials due to their size-tunable properties, yet their three-dimensional (3D) quantum confinement makes electronic states highly sensitive to structural and chemical heterogeneity, which critically impacts their optoelectronic performance. Accurately resolving the 3D atomic structure with sub-angstrom precision is thus essential for rational design. Here, we applied atomic electron tomography (AET) to determine, for the first time, the 3D atomic structure of complex core/shell QDs, resolving over 14,000 atoms per particle. Our reconstructions reveal surface morphology, eccentric cores, and nearly atomically abrupt heterovalent interfaces and identify anisotropic shell growth directed by twin boundaries. Utilizing an AET-derived atomic structure, we**

**performed large-scale quantum mechanical calculations to uncover an orientation-dependent strain accommodation mechanism where the heterogeneous strain is compensated at interfaces and twin boundaries. Furthermore, our results reveal strain-induced localized states near the band edge, which contribute to the key features of the experimental ensemble absorption spectrum. This work sets a new benchmark for atomic-level characterization, establishing a powerful framework for the rational design of next-generation nanomaterials.**

Colloidal quantum dots (QDs) were recognized with the 2023 Nobel Prize in Chemistry for their discovery and synthesis[1–3]. When the size of a semiconductor nanocrystal approaches or becomes smaller than its exciton Bohr diameter, quantum confinement substantially alters the electronic states of confined electrons or holes, leading to discrete, atom-like energy levels and spatially localized wavefunctions[4]. These quantum properties, which depend strongly on QD size, shape, and composition, can be precisely tuned through chemical synthesis, enabling a wide range of applications including light-emitting devices (LEDs)[5], solar cells[6], bioimaging[7,8] and quantum technologies[9,10]. Beyond control of size and composition, further tuning of QDs has been achieved by structural engineering such as forming complex core/shell heterostructures to suppress nonradiative recombination or introducing dopants to modulate electrical properties[11]. Beyond size and composition, atomic-scale features such as interfacial chemistry, strain, and defects strongly affect potential profiles, carrier localization, and exciton dynamics[12–15]. However, conventional transmission electron microscopy (TEM), which provides only 2D projections of 3D objects, lacks the depth resolution needed to fully reveal local crystallinity in 3D. Likewise, ensemble techniques, such as X-ray diffraction in physical science[16,17] or single particle analysis (SPA) in biological science[18,19] yield averaged structural information and are thus incapable of resolving particle-specific features like elemental exchanges at the interface or local structural distortions. The absence of comprehensive 3D structural information at the single-atomic level remains a key challenge for fully understanding and optimizing the properties of colloidal QDs.

To overcome this challenge, we employ atomic electron tomography (AET)[20–22] to directly determine the 3D atomic structure of individual InAs/ZnSe heterovalent QDs at sub-angstrom resolution. This approach localizes each atom with elemental specificity. The surface and interface orientations are also analyzed in each QD to reveal the growth process from core to shell. We find anisotropic lattice change across the core/shell interface, which arise from the overall QD morphology, twin boundaries, and lattice mismatch between InAs and ZnSe. The InAs/ZnSe interfaces were identified to be atomically abrupt while limited site-specific cation exchanges are also observed, which are highly conducted by twin boundaries and polar crystal facets such as $\{111\}$ and $\{100\}$. Taking the experimental coordinates by AET as direct input for large-scale quantum mechanical calculations, we mapped the strain field landscape across entire core/shell QDs. We uncovered a highly orientation-dependent strain accommodation mechanism that departs fundamentally from the hydrostatic paradigm typically associated with concentric core/shell QDs. Furthermore, by utilizing the atomic structure determined by AET, we identified a unique electronic structure that differs from those derived from relaxed or truncated bulk coordinates, which captures the essential features of the experimental ensemble absorption spectrum.

**3D atomic structures of complex core/shell QDs determined by AET**

Here we focus on InAs/ZnSe core/shell QDs as a representative heterovalent III–V/II–VI system, which are of particular interest for infrared LEDs[23,24], biological imaging[25], and quantum communication[26]. III–V QDs (e.g., InP and InAs) are free of toxic heavy metals such as Cd, Pb, or Hg, and offer tunable optical properties that span from the visible to the short-wave infrared region[27,28]. Due to limited shell material options, wide-bandgap II–VI semiconductors are typically used to form type-I core/shell structures. Compared to homovalent systems, heterovalent III–V/II–VI QDs exhibit more complex structures due to strong covalent bonding, large lattice mismatch, and charge discontinuities at the interface[12,29,30]. These structural complexities can critically influence their electronic and optical behavior. While considerable progress has been made in understanding their interfacial chemistry and oxidation, as well as their effects on shell growth and optical properties[31–36], direct determination of the 3D atomic structure of heterovalent core/shell QDs remains essential for investigating how atomic-scale structural heterogeneities influence QD properties. Therefore, InAs/ZnSe QDs serve as a relevant platform for exploring these structure-property relationships.

High-quality InAs/ZnSe QDs consisting of a tetrahedral ZnSe shell on a quasi-spherical InAs core were synthesized using a recently developed two-step protocol with slight modifications (Supplementary Fig. 1)[23]. The resulting InAs/ZnSe QDs exhibit a near-unity photoluminescence quantum yield (PLQY) in the near-infrared region (Supplementary Fig. 2). X-ray photoelectron spectroscopy (XPS) measurements on the InAs cores and thin-shell InAs/ZnSe intermediates showed no detectable arsenic oxide species[35]. Furthermore, electron energy loss spectroscopy (EELS) of the InAs/ZnSe QDs indicated that oxygen was concentrated primarily at the outer surface in association with oleate ligands, whereas it was negligible at the core/shell interface. Together, these results confirm the formation of an oxide-free interface under the strictly inert synthetic conditions (Supplementary Figs. 3–6). AET experiments were performed on two

InAs/ZnSe QDs (named InAs/ZnSe-1 and InAs/ZnSe-2) and one InAs QDs (named InAs-1) using annular dark-field scanning transmission electron microscope (ADF-STEM) mode (Figs. 1a, b, Supplementary Figs. 7–9, Supplementary Video 1, and Supplementary Table 1). To minimize beam damage, a low-dose data acquisition scheme[22] was used to collect datasets. After preprocessing of electron microscopic tilt series, the 3D reconstruction volume of QDs that preserves atomic-number contrast was achieved using RESIRE, a real space based iterative algorithm (Fig. 1c, Supplementary Figs. 10–12, and Supplementary Video 2)[37]. To reduce artifacts during atomic tracing and classification, we applied a newly developed universal automated local-structure-guided tracing (ALT) algorithm for complicated zinc-blende and wurtzite structures (Supplementary Figs. 13–15). The 3D atomic coordinates and chemical species of individual atoms were then determined. The total number of atoms identified in the two InAs/ZnSe QDs is 14,709 and 15,345 with a root-mean-square deviation (RMSD) of 14.05 pm (Supplementary Fig. 16). Due to the contrast limitations of HAADF-STEM, surface ligands cannot be explicitly resolved in the AET reconstruction. The present analysis therefore focuses on the inorganic InAs/ZnSe core/shell framework, from which the structural features discussed here are primarily derived.

Figures 1d–e, Supplementary Figs. 17–19 and Supplementary Video 3 show the 3D atomic configurations, compositional distributions, surface and interface morphologies of the two core/shell QDs. In both InAs/ZnSe-1 and InAs/ZnSe-2, six and seven $In^{3+}$ ions respectively, are located beyond the InAs core, as a result of the two-step synthesis method (Fig. 1e). Surprisingly, despite a 6.45% lattice mismatch between bulk InAs (6.058 Å) and ZnSe (5.667 Å), both QDs adopt a coherent single crystalline zinc-blende structure. The ZnSe shells exhibit a flattened tetrahedral geometry with truncated

edges, exposing one {111}, three {110} and one truncated {100} facet (Fig. 1f and Supplementary Fig. 20a). The InAs cores in both QDs exhibit eccentric rhombicuboctahedral morphology, with overall diameter of 3.38 nm for InAs/ZnSe-1 and 3.06 nm for InAs/ZnSe-2 (Fig. 1g, Supplementary Figs. 20b, 21 and 22). The alternating six {111}, twelve {110} and eight {100} facets of the inner InAs core's surface align well with the corresponding equal-index facets of the ZnSe outer surface. The eccentric InAs cores are positioned closer to one of the {111} facets of the outer surface, with distances 1.51 nm and 1.60 nm for InAs/ZnSe-1 and InAs/ZnSe-2, respectively. The nearest neighbours of each atom were then quantified by fitting the valley positions of the pair distribution function (PDF) and compared them with reference zinc-blende and wurtzite lattices to calculate the twinning order parameter (Supplementary Fig. 23). We find that each InAs/ZnSe QD contains at least one wurtzite-like twin boundary extending from the InAs core into the ZnSe shell (Fig. 1h and Supplementary Fig. 24). Twin boundaries oriented along the ⟨111⟩ direction are commonly low-energy features in both II–VI and III–V QDs[38,39]. In both core/shell QDs, the twin boundaries are parallel to one of the {111} facets on both tetrahedral outer surface and rhombicuboctahedral inner surface. The distance from twin boundaries to top and bottom of inner surfaces of InAs cores along perpendicular [111] directions are 0/2.86 nm and 1.50/1.14 nm for InAs/ZnSe-1 and 2.04/0.36 nm for InAs/ZnSe-2 respectively, indicating an asymmetric feature of twin boundaries in sphere-like InAs cores (Supplementary Fig. 25). Their presence disrupts the intrinsic symmetry of the InAs cores, generating polar surfaces on the top and bottom and eventually giving rise to the observed eccentric QD geometry[40–42]. Above conclusions are drawn from AET-resolved analysis of individual InAs/ZnSe QDs and therefore primarily describe particle-specific structural behavior rather than universal features of all such heterostructures. Therefore, we performed electron tomography on a cluster of InAs/ZnSe QDs at low magnification (Supplementary Figs. 26 and 27). All eight QDs within the cluster exhibit a consistent tetrahedral shape with eccentric core and twinning, confirming the representativeness of these structural properties in the sample.

**Mapping 3D local lattice change and layer curvature in core/shell QDs**

Next, we quantified the local lattice change of the QDs by comparing each atom and its nearest neighbors with a reference lattice. Because the QDs are single crystalline with several twin boundaries, the averaged lattice parameter of the whole QDs was used to determine the lattice change in each orientation. Figures 2a–c, Supplementary Figs. 28 and 29 present the layered structures and in-plane and out-of-plane components of the local lattice change for InAs/ZnSe-1, InAs/ZnSe-2 and InAs-1. The normal to the twin boundaries is defined as the z direction ($[111]$), with x and y along $[1\bar{1}0]$ and $[11\bar{2}]$ respectively. The in-plane lattice change varies smoothly and is well correlated with the core/shell structure, ranging from –8.07% (compressive) to +9.93% (tensile) (Fig. 2b). In contrast, the out-of-plane lattice change shows greater spatial heterogeneity and a wider range, from –16.92% to +17.28% (Fig. 2c). The standard deviation of the in-plane and out-of-plane lattice for the two InAs/ZnSe QDs are 1.29% and 5.54% respectively (Figs. 2e, f, and Supplementary Figs. 28e, f). The peak position of the lattice constant of the InAs core in the InAs/ZnSe QDs is 5.92 Å, which is 2.22% smaller than bulk InAs and 0.92% smaller than InAs-1. Quantitative analysis shows that lattice perpendicular to the $(111)$ plane is significantly more heterogeneous than that within the $(111)$ plane. In the other three $[111]$ directions, the in-plane and out-of-plane lattice are more isotropic (Supplementary Figs. 30–32). These results indicate that the anisotropic lattice change arises from the twin boundaries and the lattice of InAs cores in InAs/ZnSe QDs is more fluctuated than that of InAs-1 (Supplementary Fig. 29).

Figures 2g, h, and Supplementary Figs. 28g, h display the changes of lattice in and out of the (111) plane with distance from the center of InAs core. The in-plane lattice change remains stable near the core center. As it approaches the interface of InAs/ZnSe QDs, the lattice gradually decreases across the interface between the InAs core and ZnSe shell and then stabilizes with a narrow distribution across the core and shell (Fig. 2g and Supplementary Fig. 28g). The out-of-plane lattice change follows a similar overall trend, yet it exhibits markedly greater heterogeneity (Fig. 2h and Supplementary Fig. 28h). Quantitative analysis indicates that the [111] direction defined by the twin boundaries in InAs/ZnSe QDs does not interrupt the zinc-blende structure of either InAs core or the ZnSe shell, both of which remain single-phase zinc-blende structure. However, the substantial lattice mismatch between InAs and ZnSe cannot be fully resolved by the local lattice change. Our analysis indicates that the mismatch induces large fluctuations of out-of-plane lattice along [111] direction as well as global bending of the (111) planes, which is quantified using the generalized Gaussian curvature calculation method[43]. Figure 2d and Supplementary Fig. 28d show the curvature distribution of each layer in InAs/ZnSe QDs, interestingly revealing that the curvature distribution does not match the core morphology. The concaveness starts near the boundary of the InAs core and propagates only toward the thinner ZnSe shell direction that is closer to the ZnSe surface. Curves in Supplementary Fig. 33 presents the curvature profile along the [111] direction, with the InAs core boundary marked. We observe that the top ZnSe layers exhibit larger curvature and enlarged lattice even than the InAs layers, indicating asymmetric bending of InAs/ZnSe QDs towards the thin shell direction. These results demonstrate that during the shell growth, the presence of twin boundaries introduces fluctuations along the c-axis, resulting in large out-of-plane lattice change while maintaining in-plane lattice coherence. The twin boundaries also promote selection of the polar [111] direction, where one side grows more rapidly. On the opposite side, the large lattice mismatch induces bending, reducing the growth rate and leading to the formation of the tetrahedral shell structure[41,44].

**Determining the strain accommodation of InAs/ZnSe core/shell QDs**

To evaluate the energetics and stability of AET-resolved InAs/ZnSe core/shell QDs, we took InAs/ZnSe-1 sample as a representative system, and performed large-scale geometric optimizations employing the density functional tight binding (DFTB) framework (Supplementary Fig. 34)[45]. The DFTB-relaxed QD remains highly consistent with the AET-resolved structure in its global morphology and core/shell architecture (Supplementary Fig. 34a). The relaxation is accompanied by narrower bond-length distributions (Supplementary Fig. 34b) and reduced lattice distortions. The small RMSD between the AET-resolved and DFTB-relaxed structures indicates relaxation introduces only limited atomic displacements (Supplementary Fig. 34c), further supporting the high fidelity of the AET-resolved atomic structures. The AET-resolved structure possesses a higher relative energy (~0.78 eV/atom) than the fully relaxed configuration. This discrepancy highlights the distinction between the experimental core/shell QDs and its idealized computational counterpart. The former reflects the core/shell morphology formed under experimental conditions, whereas the latter converges toward the intrinsic potential-energy minimum of the system in vacuum at 0 K.

The DFTB-relaxed configuration provides a reliable reference for the AET-resolved structure, enabling a rigorous assessment of strain accommodation mechanisms across the entire core/shell QD. By mapping the experimental atomic coordinates against this DFTB-relaxed reference, we quantified the local 3D strain fields (Fig. 3 and Supplementary Figs. 35–38). We found that the strain accumulation across the core/shell QD is highly orientation-dependent. The in-plane strain is largely confined to the interfacial region and changes sign across the core/shell boundary (Fig. 3b and Supplementary Figs. 35b, e). Specifically, the ZnSe shell undergoes expansion while the InAs core experiences compression relative to the DFTB-relaxed structure. The sign change of the in-plane strain across the core/shell boundary identifies the core/shell interface as a localized strain-compensation zone where lattice-mismatch strain is redistributed under the lateral constraint of the ZnSe shell, effectively enhancing interfacial coherence. Notably, the out-of-plane strain component along the $[111]$ crystallographic direction is significantly larger than the in-plane strains and exhibits a sign inversion across the twin-boundary planes (Fig. 3c and Supplementary Figs. 35c, f). This suggests that these structural defects actively participate in local strain redistribution. By flipping the strain state, the twin boundaries help relieve the axial stress accumulation that would otherwise be unsustainable within this highly eccentric architecture. Additionally, we calculated the shear strain components in the QD, which provide further evidence for the anisotropic nature of the strain field (Supplementary Fig. 36). Compared with the relaxed structure, the AET-resolved model exhibits smaller shear strain throughout the particle. The summed shear-strain magnitude further shows that this reduction is present along all directions (Fig. 3d and Supplementary Fig. 37). We also calculated the curvature of each $(111)$ plane and found that the curvatures of the relaxed structure are larger than the corresponding ones in the AET model, particularly near the surface region adjacent to the eccentric core (Supplementary Fig. 38).

Collectively, these findings represent a highly orientation-dependent strain accommodation mechanism that departs fundamentally from the hydrostatic paradigm typically associated with concentric core/shell QDs. These results suggest that the experimentally resolved atomic coordinates accommodate interfacial lattice mismatch more effectively through localized in-plane strain, which forms a strain-compensation zone at the interface. Simultaneously, the much larger out of plane strain is relieved by twin-boundary planes that invert the strain state and reduce shear distortion within the eccentric structure.

**Unveiling heterovalent interfaces between InAs core and ZnSe shell**

For II–VI/III–V core/shell QDs, variation in dipole densities arising from interfacial heterovalency has been shown to strongly influence the electrostatic potential distribution and band alignment[12,30]. Accurately determined 3D atomic structures of the QDs enable direct characterization of the core/shell interface. Contrary to the expected gradient interface[30], we find the as-synthesized QDs exhibit atomically sharp interfaces, where the concentration of $In^{3+}$ decreases from 86.34% to 4.09% in IA InAs/ZnSe-1 and from 84.28% to 5.14% in InAs/ZnSe-2 within one single atomic layer (Figs. 4a, b, and Supplementary Figs. 39a, b). At the InAs/ZnSe interfaces, we observe 52 cations exchanges (between $In^{3+}$ and $Zn^{2+}$) for InAs/ZnSe-1 and 45 for InAs/ZnSe-2 (Fig. 4c and Supplementary Fig. 39c). The excess $Zn^{2+}$ cations (33 in InAs/ZnSe-1 and 26 in InAs/ZnSe-2) are likely due to the leaching of exchanged $In^{3+}$ cations during the second synthesis step[23]. Mapping the orientations of the exchanged cations on stereographic projections reveals a strong preference for {111} and {100} facets (blue and red regions), indicating facet-dependent chemical reactivity during interdiffusion (Figs. 4c, d, and Supplementary Figs. 39c, d). We then counted the number of exchanged cations on each (111) layer and found that $Zn^{2+}$ substitution occurred most frequently near the twin boundary: 66.67% for InAs/ZnSe-1 and 42.31% for InAs/ZnSe-2, respectively (Fig. 4e and Supplementary Fig. 39e). The continuous symmetry measures (CSM) were then applied to quantify tetrahedral distortion[46,47]. Elevated CSM values, showing pronounced distortion, correlate with the large number of cation exchanges near the twin boundary (Fig. 4f and Supplementary Fig. 39f).

For zinc-blende and wurtzite structures, each anion/cation has four nearest neighbors. Upon cation exchange, heterovalent bonds form at the interface, inducing variable local strain and electric fields. At the interface, 556 heterovalent bonds are identified, including 287 positive (In–Se) and 269 negative (Zn–As) bonds (Supplementary Fig. 40). Radial analysis of the average CSM values from the center of InAs core indicates low CSM values of 0.31 ± 0.02 within the InAs core, but a sharp increase at the interface, followed by a subsequent decrease towards the surface (Supplementary Fig. 41). Local dipoles, derived from asymmetric atomic displacements relative to ideal lattice positions, are enhanced by these heterovalent bonds. 3D visualization of the dipole vectors reveals preferential alignment compared with the interface and crystal facets (Figs. 4g, h, and Supplementary Fig. 42). Statistical analysis indicates that the local dipole magnitudes are larger at the {111} and {100} facets, consistent with their polar nature (Fig. 4i). Radial mapping from the InAs core to ZnSe shell reveals a gradual emergence of dipole moments across the core/shell interface (Fig. 4j). Notably, dipole magnitudes are enhanced near twin boundaries as well, in agreement with the earlier finding that these regions are structurally distorted and chemically active (Fig. 4k). Together, these results suggest that local dipoles, reflecting both interfacial strain and symmetry breaking, may play a critical role in directing cation exchange, particularly along specific crystallographic planes and near planar defects such as twin boundaries.

**Correlating 3D atomic structure with the electronic and optical properties of colloidal QDs**

To examine how the experimentally resolved atomic structure influences the electronic and optical properties of QDs, we performed DFT calculations using atomic coordinates of the InAs core in both InAs/ZnSe-1 and InAs/ZnSe-2, directly obtained from AET (Fig. 5 and Supplementary Figs. 43–49). The type-I band alignment of InAs/ZnSe localizes both electron and hole predominantly within the InAs core, which enables us to retain realistic internal features and heterogeneous shell-induced strain within the InAs core while reducing the system to a computationally accessible size. We therefore extracted the InAs core directly from the AET-resolved structure and passivated the surface dangling bonds. We refer to this as the experimental QD (exp. QD). To assess the effect of shell-induced strain, we generated a relaxed model (relaxed QD) via full geometry optimization of the exp. QD. While the exp. QD inherits the heterogeneous strain of the parent ZnSe shell yet is liberated from its geometric constraints, the relaxed QD represents an energy-minimized state in which these constraints are absent. Finally, a nearly spherical QD truncated from a zinc-blende InAs bulk crystal was used as a reference model (ref. QD). Because this reference model neglects shape anisotropy, strain variations, and twin boundaries, the comparison among these models highlights the importance of the AET-resolved realistic structures for capturing the main experimental optical features (Supplementary Figs. 43 and 44).

The experimental QD was extracted from the InAs/ZnSe-1 sample for demonstration. In the exp. QD, heterogeneous strain widens the electronic bandgap by ~

0.57 eV compared to the relaxed QD (Fig. 5a). It also introduces energetic sparsity, resulting in lower electronic density of states (DOS) (Fig. 5a). Moreover, the local lattice distortion induced by the heterogeneous strain promotes localization of the band-edge electronic states. This effect is supported by the calculated inverse participation ratios (IPRs), where low IPRs indicates localization, and by the selected plots of frontier molecular orbital states (Figs. 5b–d). Compared to the relaxed QD, the highest occupied molecular orbital (HOMO) state in the exp. QD exhibits pronounced charge localization around the twin boundary, corresponding to a lower IPR (Fig. 5d). Notably, higher unoccupied states (e.g., LUMO+1) and deeper occupied states (e.g., HOMO-1 and HOMO-2) also show significant charge localization near the surface (i.e., surface-like states), with nearly vanishing IPRs (Figs. 5 b–d and Supplementary Figs. 45–47). In contrast, both the relaxed and ref. QDs display extended, bulk-like electronic states, and the twin boundary alone does not induce charge localization, as indicated by the corresponding orbital plots and higher IPRs.

The calculated optical absorption spectrum of the exp. QD captures the key features of the experimentally measured ensemble spectrum, including both the energy position of the first pronounced excitonic peak and the overall lineshape (Fig. 5e)**.** Similar results have been obtained for exp. QD extracted from InAs/ZnSe-2 (Supplementary Fig. 48). The first excitonic peak is blue shifted by up to ~ 0.56 eV relative to the relaxed InAs QD. The distinctive spectral profile of the exp. QD is found to strongly correlate with the localization of the electronic states. The first bright excitonic state arises exclusively from the HOMO–LUMO transition, in which the two orbitals exhibit strong spatial overlap, leading to a large oscillator strength (Figs. 5e–h). In contrast, higher excitonic states between the first and second pronounced peaks are surface-to-bulk states in nature, involving transitions from deeper, surface-localized occupied states (e.g., HOMO-1 in Fig. 5b) to the bulk-like LUMO states (Fig. 5h and Supplementary Fig. 49). These exciton states, characterized by weak oscillator strengths and long radiative lifetimes, are effectively optically dark and account for the dip-like feature observed near the first excitonic peak in the experimentally measured absorption spectra (Fig. 5e).

The density of transitions in the exp. QD appears much lower than in the relaxed and ref. QD, particularly in the lower-energy region of the absorption spectra (Figs. 5e–h), correlating with the reduced DOS in the exp. QD (Fig. 5a). Neither the energy position of the first excitonic peak nor the overall lineshape of the experimentally measured absorption spectrum is captured by the relaxed or ref. QD, as their excitonic states involve only bulk-like transitions (Figs. 5f–h). To assess the impact of the twin boundary on optical transitions, we compared the calculated absorption spectrum of the relaxed QD with that of the ref. QD. While their absorption edges are comparable in energy, their spectral shapes differ significantly, due to the impacts of QD shape anisotropy and twin boundary. Notably, the twin boundary in the relaxed QD causes a ground-state exciton with minimal oscillator strength ($\sim 10^{-3}$), which is thus optically dark (Fig. 5f). Although this exciton originates from a nominally allowed HOMO-LUMO transition, its optical inactivity results from the non-trivial interplay between symmetry mixing and confinement effects on the valence band structure. In the exp. QD, heterogeneous strain breaks down the symmetry effects, thus further violating this optically forbidden character (Fig. 5e). These results highlight the critical role of precise 3D atomic structures of QDs, as resolved by AET, in accurately describing their properties. Incorporating local features such as strain, symmetry distortion, and internal boundaries not only enables agreement with experimental measurements, but more importantly, provides deeper insight into the properties of QDs.

**Conclusions**

After decades of development and successful commercialization in areas such as display technologies, colloidal QDs, whose optoelectronic properties are highly sensitive to subtle structural features, have reached a stage where atomic-resolution characterization in 3D is essential to fully elucidate structure-property relationships and enable further performance improvements. Here we address this gap by newly developed AET method with the ALT algorithm to resolve the 3D atomic structures of high-quality InAs/ZnSe core/shell QDs, which clarify the relationship between buried structural defects and QDs growth mechanisms. Using the experimental coordinates as direct input for large-scale quantum mechanical calculations, we reveal an orientation-dependent strain accommodation mechanism. The in-plane strain is concentrated at the core/shell interface, where the ZnSe shell expands and the InAs core compresses to create a strain-compensation zone. Meanwhile, the much larger out-of-plane strain along the [111] direction is redistributed though twin-boundary planes. The enhanced hydrostatic strain and reduced shear strain both contribute to the improvement of interfacial coherence. Furthermore, the single-crystal structure preserves an abrupt InAs/ZnSe interface. Limited cation exchanges occur preferentially along polar crystal orientations and near the twin boundaries, consistent with the direction of the interfacial dipoles. We also identify pronounced, heterogeneous strain effects on the electronic structure, including particularly the existence of highly localized states near the band edge. These electronic features enable us to capture the essential features of the experimentally measured ensemble absorption spectrum. These results show that AET can resolve 3D atomic structures of complex QDs, quantify strain and interfacial chemistry, and establish a significant link between theoretical analysis with optical measurements.

Previous applications of AET primarily focused on resolving 3D atomic structures in close-packed metallic systems including both crystalline and amorphous. With the capability of newly developed AET method with the ALT algorithm, this work is, to our knowledge, the first substantive extension of AET to complex semiconductor heterostructures, enabling quantitative 3D mapping of strain, composition, and defects in non-close-packed structures. Establishing such atomic-scale metrology for semiconductor nanomaterials is crucial for engineering chips and integrated photonic components including metal-semiconductor junctions and hybrid nanoarchitectures, thus opening new pathways for advancing semiconductor technology.

**Figures and Figure legends**

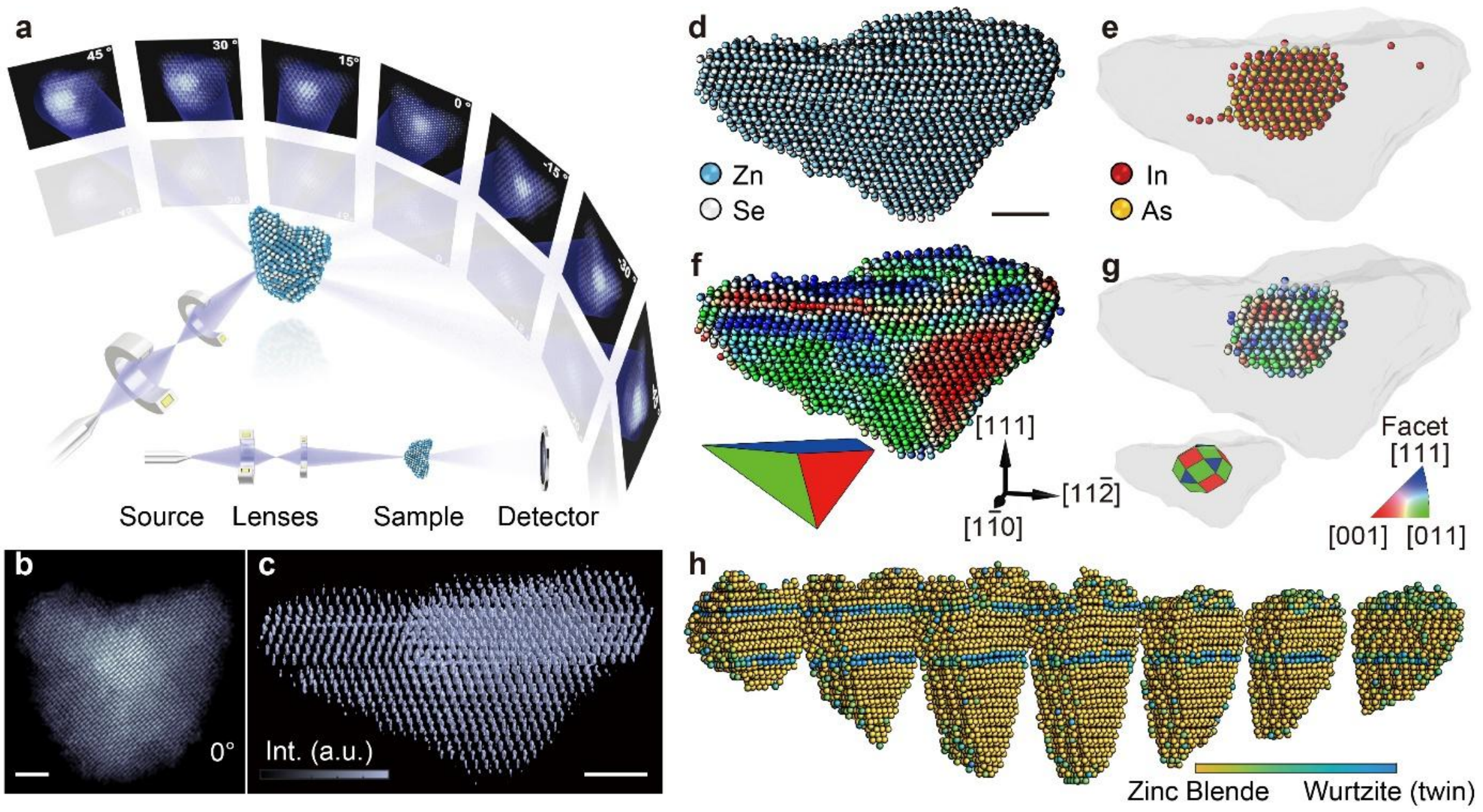

**Fig. 1 | 3D atomic structure of a complex InAs/ZnSe core/shell QD determined by atomic electron tomography (AET). a**, Schematic of the data acquisition procedure of AET. **b**, Experimental projections InAs/ZnSe-1 at 0°. **c**, 3D reconstruction volume of InAs/ZnSe-1 with contrast representing atomic number ($In^{3+}$ different from $As^{3-}$, $Zn^{2+}$ and $Se^{2-}$). **d-e**, 3D atomic structures traced from the reconstruction in (**c**). **f**, Surface facet orientation of InAs/ZnSe-1. **g**, Interface orientation of InAs/ZnSe-1, exhibiting a rhombicuboctahedron shape. **h**, Twin order parameters of InAs/ZnSe-1 with a slices view. Scale bar, 2 nm.

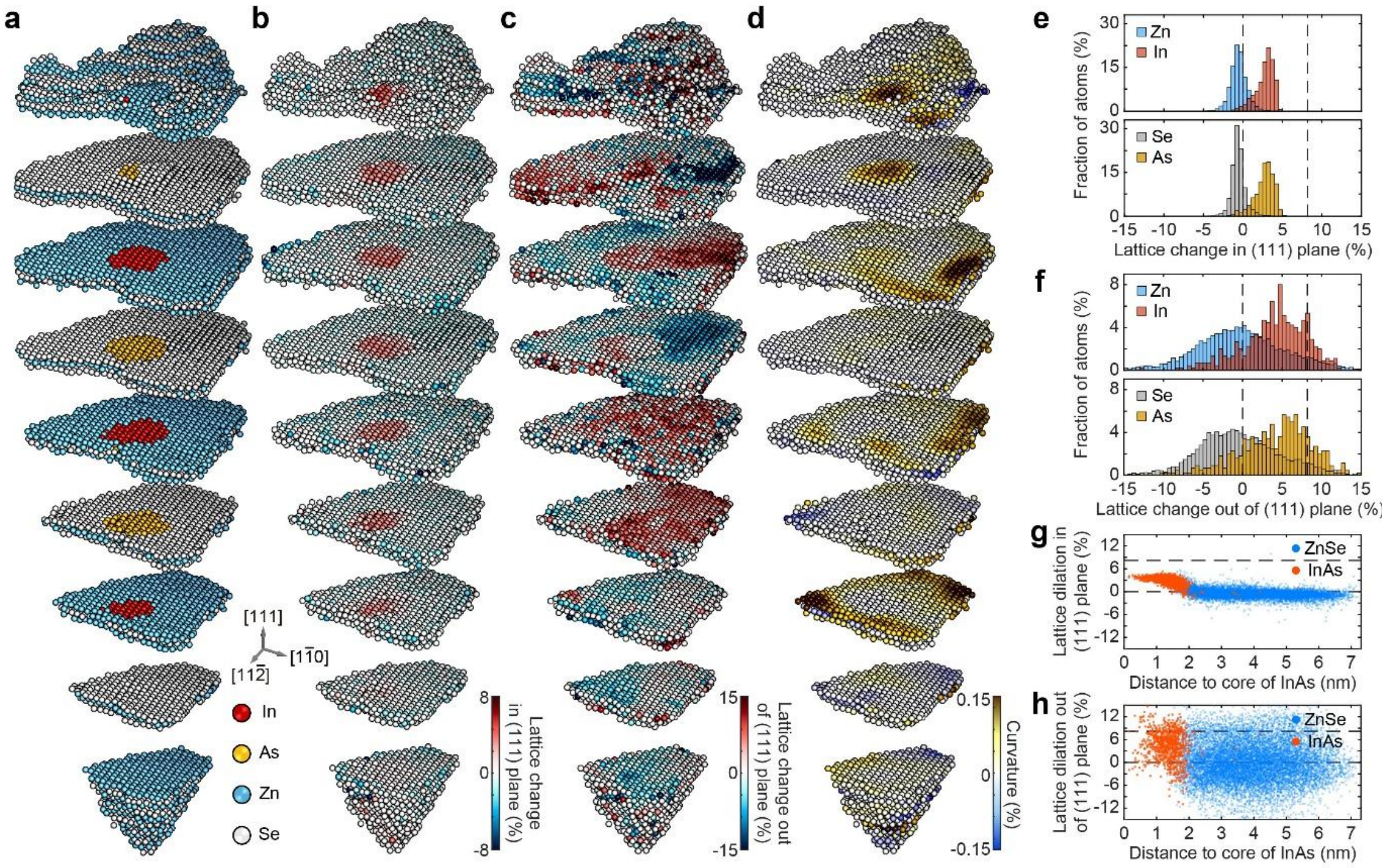

**Fig. 2 | 3D lattice change and curvature analysis of InAs/ZnSe-1 QDs. a-d**, Layered 3D atomic model (**a**), in-plane lattice change (**b**), out-of-plane lattice change (**c**) and layered curvature (**d**) in InAs/ZnSe-1 along the [111] direction. **e**, Statistical distribution of in-plane lattice change of In, Zn, Se and As atoms. **f**, Statistical distribution of out-of-plane change of In, Zn, Se and As atoms. **g-h**, Lattice distortion distribution in and out of (111) plane of InAs/ZnSe-1 from the InAs core center to the ZnSe shell.

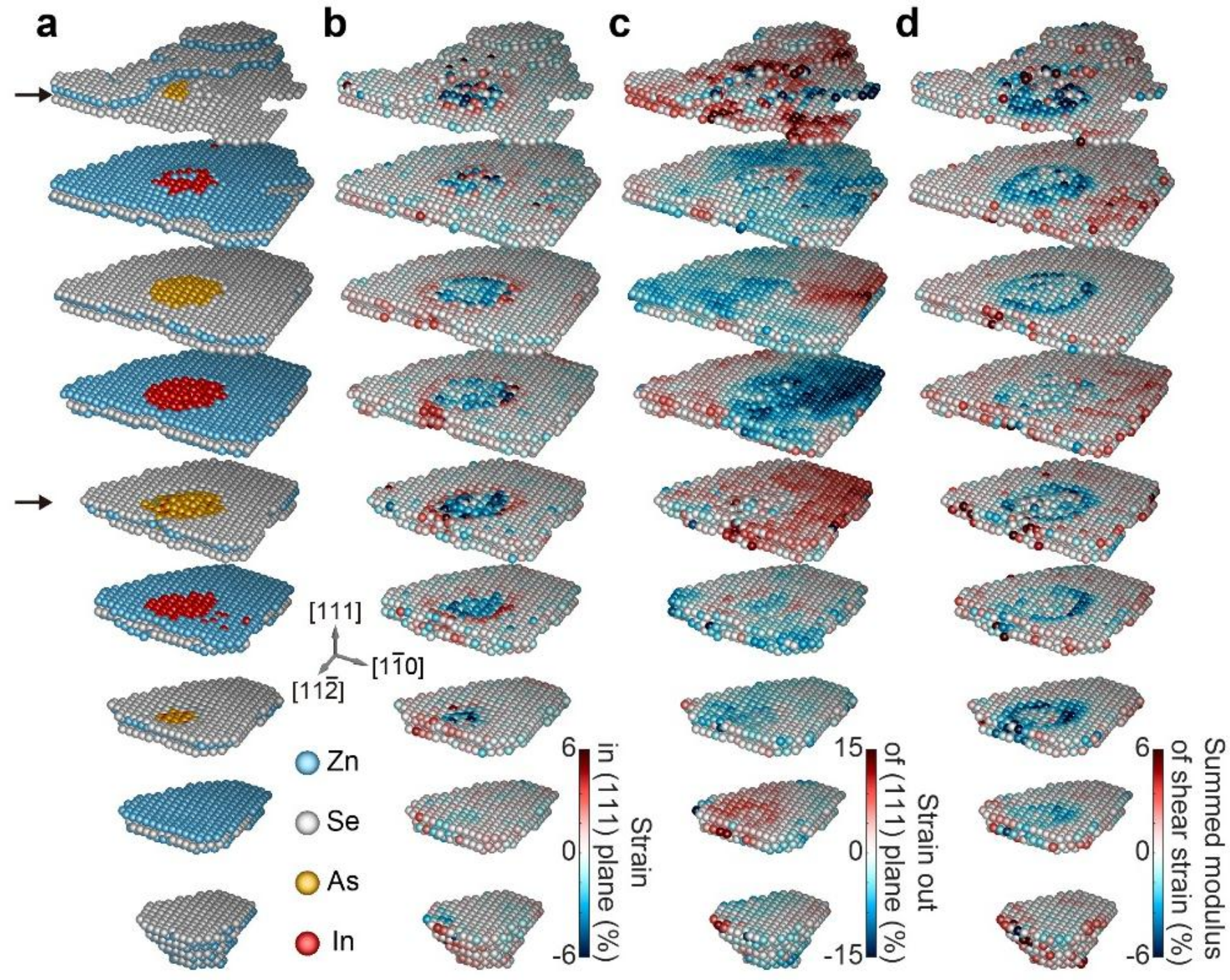

**Fig. 3 | Strain accommodation of InAs/ZnSe core/shell QDs. a-d**, Layered 3D atomic model (**a**), in-plane strain (**b**), out-of-plane strain (**c**), and summed modulus of shear strain (**d**) in InAs/ZnSe-1 along the [111] direction. The in-plane and out-of-plane strain are derived by subtracting the lattice change of DFTB-relaxed atomic model from AET-resolved experimental data. The summed modulus of shear strain is calculated by subtracting the shear strain of the DFTB-relaxed model from the corresponding AET-resolved value, where the shear strain is defined as $\sqrt{(\varepsilon_{xy})^2 + (\varepsilon_{xz})^2 + (\varepsilon_{yz})^2}$. The arrows indicate the position of twin boundaries.

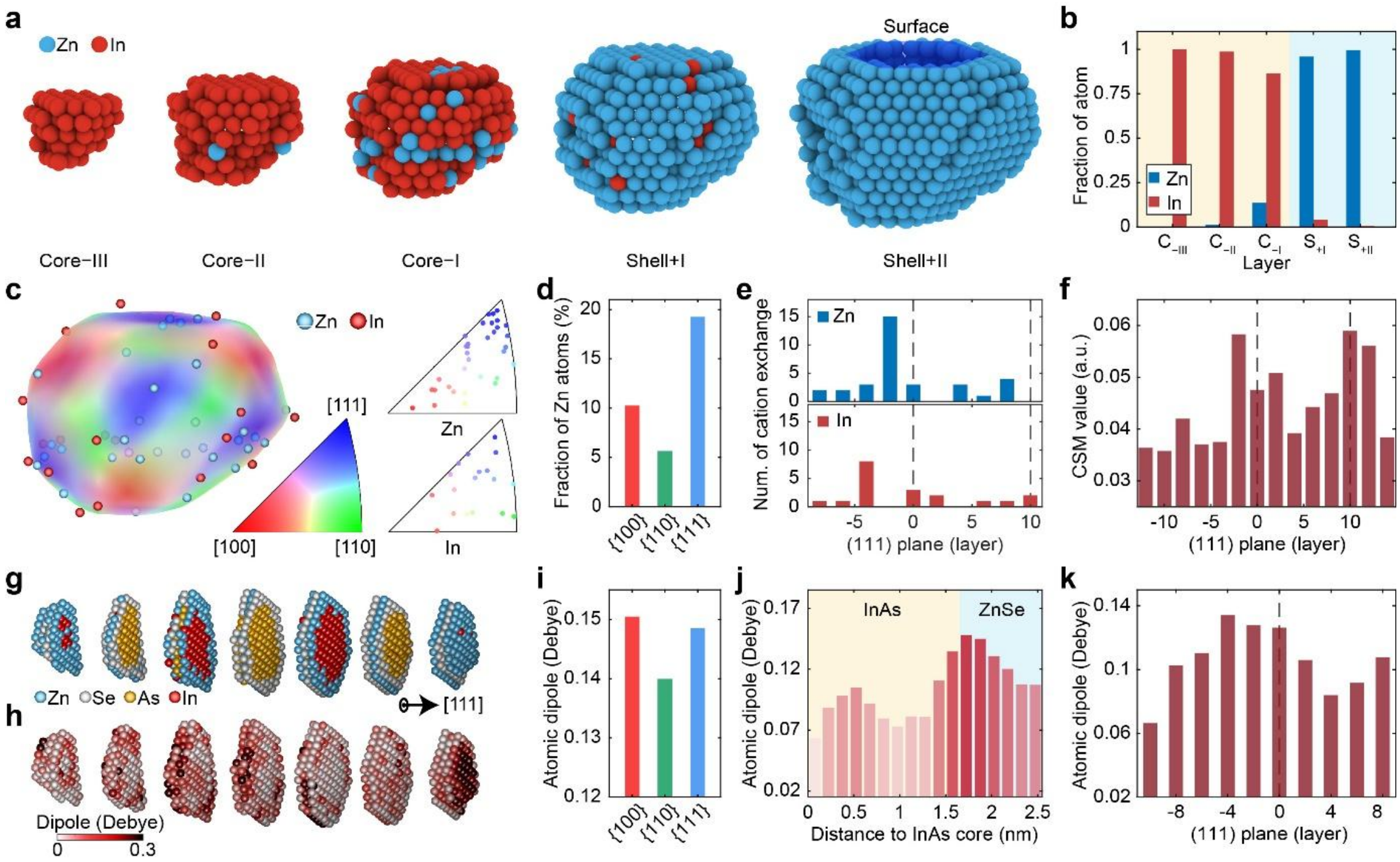

**Fig. 4 | Cation distribution, interface structure, and local symmetry breaking in the InAs/ZnSe core/shell interface at the 3D single-atom level.** **a**, Layer-by-layer distribution of $In^{3+}$ and $Zn^{2+}$ cations near the core/shell interface, extracted from the 3D reconstructed atomic model. Core–I ($C_{-I}$), Core–II ($C_{-II}$) and Core–III ($C_{-III}$) denote the first, second, and third surface cationic layers of the InAs core, respectively. Shell+I ($S_{+I}$) and Shell+II ($S_{+II}$) denote the first and second layers of shell cations positioned nearest to the InAs core, respectively. **b**, Quantitative statistics of In and Zn atomic populations as a function of radial layer. **c**, Spatial distribution of cation exchange sites. **d**, Facet-resolved statistics of cation exchange events. **e**, Histogram of cation exchange number occurring on each atomic layer along the [111] direction. The dashed line marks the position of the twin boundary with layer 26 designated as layer 0 (Supplementary Fig. 18a), serving as the reference twin position. **f**, Layer-resolved average of the continuous symmetry measure (CSM) for each (111) cationic plane. Higher CSM values correspond to larger deviations of the local tetrahedral environment from ideal symmetry. **g, h**, Sliced atomic

model (**g**) and corresponding distribution of dipole vectors (**h**) derived from atomic displacements in the reconstructed structure. **i**, Statistical analysis of the correlation between local dipole orientation and interfacial facet orientation. **j**, Radial distribution of dipole vectors from the InAs core centroid to the ZnSe shell. **k**, Relationship between local dipole magnitude and distance from the twin boundary, considering only cationic layers.

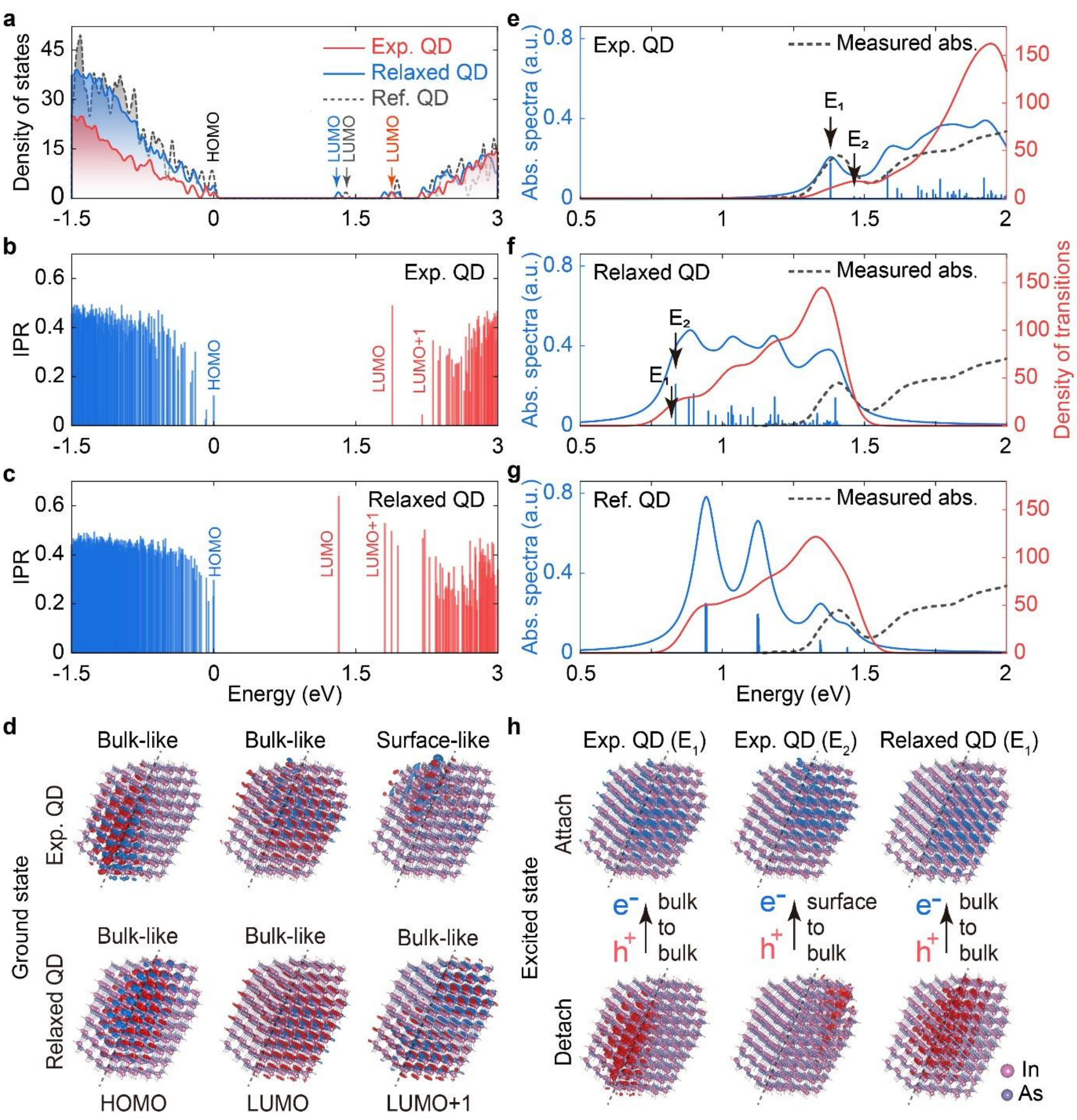

**Fig. 5 | Electronic band structures calculated from experimental coordinates of core/shell QDs. a**, Electronic density of states (DOSs) calculated using the InAs/ZnSe-1 experimental coordinates (Exp. QD), relaxed coordinates (i.e., relax the experimental coordinates to their equilibrium configuration, which was referred as relaxed QD), and reference QD (Ref. QD) which was cut from bulk InAs with zinc-blende structure while assuming a nearly spherical shape. All QDs have nearly equal number of atoms. The energies of the highest occupied molecular orbital states for all those QDs were aligned, which is further considered as reference energy. **b, c**, Calculated inverse participation ratios (IPRs) of all electronic states within the energy window of −1.5 eV to 3 eV for both Exp. QD and Relaxed QD. **d**, Molecular orbital plots for selected frontier orbital states in both Exp. QD and Relaxed QD, which show either highly extended bulk-like feature or highly localized surface-like feature. The black dashed line across the plot indicates approximately the position of the grain boundary. **e-g**, Calculated optical absorption spectra and density of transitions for Exp. QD, Relaxed QD and Ref. QD, respectively, compared with the experimentally measured absorption spectra. Each absorption spectrum is generated by calculating the lowest 50 singlet excitonic states. The vertical line shows the absorption peak corresponding to each exciton state. A Lorentzian broadening function is employed with broadening parameter of 50 meV. **h**, The calculated attach and detach densities of selected exciton states in Exp. QD and Relaxed QD highlighted in (**e, f**) which correspond to either bulk-to-bulk transition or surface-to-bulk transition, respectively. The black dashed line across the plot indicates approximately the position of the grain boundary.

**Author contributions** Y. Y., B. J. and Z. Z. conceived the idea, designed the experiments, and supervised the research project. J. Z. synthesized the materials and performed the optical absorption and XPS test. Y. Y., Q. W. and T. X. conducted the AET experiments. Y. Y., Q. W., T. X. and C. O. performed the reconstruction, atom identification and structural analysis. Z. Z., M. P. and W. X. performed the large-scale materials simulations. Y.Y., B. J., Z. Z., Q. W. and C. O. wrote the paper. All authors contributed to the experiments and discussion of the manuscript.

**Competing interests** The authors declare no financial and non-financial competing interests.

## METHODS

**Materials**

Zinc acetate ($Zn(Ac)_2$), 99.99%), indium acetate ($In(Ac)_3$, 99.99%), di-n-octylamine (DOA, 97%), and squalane (96%) were purchased from Sigma-Aldrich. Selenium (99.99%) and trioctylamine (TOA) and indium fluoride ($InF_3$, 99.99%) were purchased from Macklin. Zinc fluoride ($ZnF_2$, 99%) and 1-octadecene (ODE, 90%) were purchased from J&K Scientific. Oleylamine (OAm, 80%-90%) was purchased from Strem Chemicals. Oleic acid (OA, 99%) was purchased from Psaitong. Tris(trimethylsilyl)arsine ($(TMS)_3As$) was synthesized in our lab following a well-established procedure. The solvents of TOA, ODE and squalane were degassed at 110 °C for 1 hour and stored in an $N_2$-filled glovebox.

**Preparation of precursors**

InAs clusters: 1.5 mmol of $In(Ac)_3$, 0.75 mmol of $InF_3$, 4.5 mmol of OA, and 4.5 mL of squalane were mixed in a three-neck flask and degassed under vacuum at 120 °C for 1 hour. Afterward, $N_2$ was introduced, and the mixture was allowed to cool naturally to room temperature. Then, a solution containing 0.75 mmol of $TMS_3As$, 2.25 mmol of DOA, and 1.5 mL of squalane were added to the flask under vigorous stirring to prepare the InAs clusters.

$ZnOA_2$: 1.47 g of $Zn(Ac)_2$, 5.15 mL of OA, and 14.85 mL of TOA were loaded into a three-neck flask and degassed under vacuum at 120 °C for 1 hour to obtain 0.4 M $ZnOA_2$.

TOP-Se: 5.53 g of selenium powder was dissolved in 35 mL of TOP in a glovebox, and the mixture was diluted with TOA to obtain 0.4 M TOP-Se.

**Synthesis of InAs core**

InAs core (900 nm): A mixture of 1 mmol of $In(Ac)_3$, 0.5 mmol of $InF_3$, 3 mmol of OA, and 5 mL of ODE was loaded in a 50 mL three-neck flask and was degassed at 110 °C for 1 hour. The mixture was then heated to 300 °C under $N_2$, and a solution containing 0.5 mmol of $TMS_3As$, 1.5 mmol of DOA and 1 mL of ODE was rapidly injected into the flask and maintained for 30 min. The InAs cores were further grown by injecting InAs clusters into the flask using a syringe pump at a rate of 4.2 mL/h, until the first excitonic absorption peak reached 900 nm. To prepare InAs nanocrystals with a diameter of approximately 10 nm, a mixture of 6 mL of the InAs core solution (900 nm) and 1 mmol of $InF_3$ was added to a three-neck flask and degassed under vacuum at 110 °C for 10 minutes. Under a $N_2$ atmosphere, the mixture was heated to 300 °C, and additional InAs clusters were injected using a syringe pump. This growth process was repeated in cycles until the target particle size was achieved.

**Synthesis of InAs/ZnSe core/shell quantum dots**

6.0 mL of TOA, 0.12 g of $Zn(Str)_2$, and 20.8 mg of $ZnF_2$ were placed in a three-neck flask and degassed under vacuum at 120 °C for 30 minutes. Under $N_2$, 0.95 mL of InAs core solution was introduced, and the mixture was then heated to 340 °C. To initiate ZnSe shell growth, 1.6 mL of $ZnOA_2$ (0.4 M) and 1.5 mL of TOP-Se (0.4 M) were added dropwise in sequence, followed by incubation for 30 minutes. The ZnSe shell thickness was tuned by repeating the sequential addition of $ZnOA_2$ (0.4 M) and TOP-Se (0.4 M). Prior to each addition cycle, 0.4 mL of $ZnF_2$ solution (50 mg/mL in TOA) was added to the mixture[23]. After three cycles, the resulting products were washed with toluene and ethanol to yield purified InAs/ZnSe core/shell quantum dots.

**Optical Measurements**

The absorption spectra were recorded using a Jasco V-770 UV-vis-NIR spectrophotometer. Samples were dispersed in 3 mL of toluene in 1 cm path length quartz cuvettes with airtight screw caps in a $N_2$-filled glovebox. Steady-state and time-resolved PL measurements and PL quantum yield (PLQY) were performed using an Edinburgh FLS 1000 fluorescence spectrometer. All samples were diluted to an optical density of $0.1 \pm 0.02$ at the excitation wavelength to minimize reabsorption effects.

**AET data acquisition**

InAs and core-shell InAs/ZnSe quantum dots dispersed in hexane solution were deposited on to silicon nitrite membranes (Norcada), and were heated at 80 °C for 24 hours in vacuum to alleviate contamination. The AET experiments of QDs were performed using the aberration corrected Thermo Fisher Scientific Spectra Ultra S/TEM microscope at the Instrumentation and Service Center for Physical Sciences (ISCPS) of Westlake University, operated in ADF-STEM mode at 300 kV (Supplementary Table 1). The convergence angle is 21.44 mrad and collection angle of HAADF is 34 to 200 mrad. The pixel size of 5.10 Mx image acquisition was calibrated by (100) facet of the standard $SrTiO_3$ crystal. During the acquisition of ADF images, adjacent QDs around the region of interest were employed for optimizing various aberration and calibrating focus to prevent it from excessive electron exposure[22]. To further minimize the drift distortion and electron dose at each tilt angle, three sequential images were taken with a dwell time of 3 μs. The total electron dose of each tilt series was optimized to be between $5.20\times10^5$ and $5.68\times10^5$ $e^-/Å^2$ to reduce the beam damage (Supplementary Table 1). To ensure the structural stability of QDs under electron irradiation, zero-tilt projections were compared before, during, and after the tomographic tilt series. Slight surface damage of QDs was observed after tilt series acquisition, which was a predictable outcome of prolonged STEM imaging at atomic resolution. It mainly manifested the migration of surface atoms under electron-beam irradiation, leading to a thin disordered layer on the surface whose weak signals were likely to be averaged out in the subsequent reconstruction process[48,49]. As a result, the influence of the disordered surface atoms on the reconstructed internal atomic structure of QDs is negligible.

**Image pre-processing**

A multi-pronged image pre-processing protocol was performed on each AET dataset as outlined below.

i) Drift correction. To mitigate projection distortion at each tilt angle due to the instability of the stage, drift correction was applied to three subsequent images employing a normalized cross-correlation algorithm. The maximum cross-correction coefficients were computed to determine the sub-pixel shifts and drift rates of latter two projections relative to the first. The resulting drift was corrected by interpolating raw projections accordingly. Three corrected projections at every angle were then averaged[50].

ii) Image denoising. The drift-corrected ADF projections were yielded with a mixture of Poisson and Gaussian noise and were denoised using Block-matching and 3D filtering (BM3D), which has demonstrated robust performance in denoising raw data of AET[21,22,50]. We first estimated the parameters of Poisson and Gaussian noise in the corrected image stacks. And the BM3D parameters were optimized by quantitatively minimizing the discrepancy between the denoised projections and simulated projections, ensuring maximal structural consistency. The optimal parameter sets were then subsequently applied to the full dataset for high-fidelity denoising[52].

iii) Background subtraction. After denoising, the QDs boundaries can be identified by exploiting their contrast differences in the ADF-STEM projections and generated binary masks. Laplacian interpolation combined with these masks, enabling precise estimation and subtraction of the background intensity of projections of QDs[53].

iv) Image alignment. The alignment of the background subtracted images was achieved by the common line method along the tilt axis and the center of mass method perpendicular to the tilt axis, which have been demonstrated to align experimental tilt series with sub-pixel accuracy[22,54].

**Tomographic reconstruction**

Three-dimensional (3D) reconstruction of QDs was computed using the Real Space Iterative Reconstruction (RESIRE) algorithm[37]. In each iteration, the Radon transform was applied to generate the forward projections of the current object. The residual gradient was calculated by comparing the computed projections with the experimentally measured images. This residual gradient was then linearly back-projected into the 3D volume and used to update the object in the direction of descent. A 3D reconstruction of QDs was generated after about 200 iterations.

Subsequent angular refinement and spatial re-alignment were carried out to further minimize errors of tilt angles and projections alignment, resulting from instrument misalignment and in-plane rotation induced by electron beam. A series of 2D projections was firstly calculated from the 3D reconstruction by altering three Euler angles based on the experimental tilt angles. Comparison of the calculated and measured projections was then performed to obtain the minimum error metrics, thereby determined the real tilt angles and image offset of QDs. This procedure followed an iterative refinement process that gradually converged. A precise final 3D reconstruction can be calculated using refined angles and aligned projections based on RESIRE algorithm.

The resolution of the 3D reconstruction is limited primarily by the missing wedge imposed by the tomography geometry. In our experiment, the tilt range is restricted to approximately −70° to +70°, because higher tilts are blocked by the grid and holder. This incomplete angular coverage reduces the resolution along the depth direction. The resolution on missing wedge direction (depth direction) will be:

$$r_z = \frac{1}{\frac{1}{r_x} \times \cos\frac{\theta}{2}} = r_x \sec\frac{\theta}{2}$$

Where $r_x$ is the spatial resolution of 2D images (in-plane resolution), $\theta$ is the missing angle. For HAADF-STEM mode in our Thermo-fisher Spectra Ultra microscope, the in-plane spatial resolution is 50 pm, and our missing angle is 40° (180°−140°). So, the depth resolution of our tomography reconstruction is 53.2 pm, which remains well below the typical interatomic distance in these quantum dots (>2 Å).

**Determining the 3D atomic coordinates and species / types**

To enhance the precision of atomic positions, we first apply linear interpolation to the 3D reconstruction. Local maxima are then traced within the reconstructed volume as candidate atomic sites. A 3D polynomial fitting method is subsequently performed to determine the exact atomic positions within each local volume of 0.87 Å × 0.87 Å × 0.87 Å, with a constraint that the minimum distance between any two neighboring atoms is greater than 2 Å[55,56].

After obtaining the list of potential atomic coordinates, we performed manual corrections to address missing or erroneously identified atomic positions by systematically evaluating the correspondence between individual atoms in the preliminary atomic model and the reconstructed volume.

In the 3D volume reconstructed from ADF images, the intensity of each atom's local volume is influenced by its atomic number; specifically, the higher the atomic number, the stronger the atomic contrast. The atomic numbers of Zn, As, Se, and In are 30, 33, 34, and 49, respectively. Since the atomic numbers of Zn, As, and Se are close to each other, direct distinction of these four elements based solely on atomic local intensity is challenging. Here, we first performed Gaussian clustering on the intensities of atoms in the 3D reconstruction volume, identifying the cluster with the highest intensity as the potential In atoms. Then, based on the crystallographic arrangement rules of cations and anions on the (111) plane in the zinc blende structure, combined with the positions of the potential In atoms, we defined the cation and anion layers within the atomic model. Subsequently, Gaussian clustering was applied to all atoms in the cation layer to accurately differentiate between Zn and In atoms. Using the positions of In atoms, we can approximately locate the As atoms within the InAs core, ultimately achieving a full classification of the four elemental species (Zn, As, Se, and In) in the 3D atomic model.

**Multi-slice method for precision estimation**

To evaluate the 3D precision of the AET method, we carried out multi-slice simulations to obtain the computed ADF-STEM projections based on the atomic model of InAs/ZnSe-1[57,58]. The simulation was performed using the same experimental conditions as the tilt series acquisition (conditions listed in Supplementary Table 1). A total of 56 computed projections with refined experimental Euler angles were obtained and further convoluted with a Gaussian kernel to incorporate the influence of the probe size. Then a 3D volume was reconstructed from the computed projections employing the same reconstruction RESIRE algorithm and a new 3D atomic model was traced by the same atomic training method. By quantitatively comparing this simulated atomic model with the experimentally derived atomic model, we determined that 99.66% of the atomic positions coincide, corresponding to a root-mean-square positional deviation of 14.05 pm.

**Continuous Symmetry Measures**

The continuous symmetry measure (CSM) was exploited to quantitatively estimate the degree of symmetry of the local tetrahedral coordination structures in QDs[46,47]. The quantity can be calculated by:

$$CSM = min \frac{\sum_{k=1}^{4} |Q_k - P_k|}{\sum_{k=1}^{4} |Q_k - Q_0|}$$

Where $Q_k$ $(k = 1, 2, 3, 4)$ represents the vertex atoms, which is the four nearest neighbors, forming the local tetrahedral configuration of each atom, $Q_0$ is the center of mass of $Q_k$, and $P_k$ $(k = 1, 2, 3, 4)$ refers to the vertex of a perfectly symmetric tetrahedron. The deviation between $Q_k$ and $P_k$ was used to calculated the symmetric descriptor CSM, where CSM=0 corresponds to perfect symmetry. As the local tetrahedral geometry becomes increasingly distorted, the CSM value increases accordingly.

**Automated local-structure-guided tracing (ALT) algorithm**

To objectively determine the 3D atomic lattice from the reconstruction volume and mitigate the tracing errors of conventional methods, we developed the ALT algorithm.

i) Layer-by-layer propagation

The ALT algorithm utilizes the initial atomic coordinates derived from polynomial tracing as its model input. To establish a high-confidence origin, the atom located near the center of mass of the initial model that exhibits the maximum local intensity in reconstruction volume ($V(\boldsymbol{r})$) is selected as the propagation seed. Recognizing that both zinc-blende and wurtzite crystal phases share a tetrahedral local coordination environment, the algorithm leverages these fundamental geometric constraints to systematically propagate the atomic lattice layer-by-layer outward from the seed. To distinguish naturally under-coordinated atoms at the surface, a 3D tight support mask is generated from reconstruction volume, which preserves valid unsaturated surface atoms without forcing unphysical coordination.

ii) Geometric verification and noise pruning

For each central atom ($\boldsymbol{r}_0$), the local coordination number $n_c$ is determined using a threshold search radius of 3.40 Å. The structural fidelity of the local neighborhood is strictly evaluated to identify accurate lattice sites and prune ghost atoms. For fully coordinated sites ($n_c = 4$), the algorithm verifies the physical validity of the tetrahedron formed by the central atom and its four neighbors using bond lengths, bond angles (ideally 109°28'), and the CSM. If these geometric criteria are satisfied, the central atom is confirmed as accurate, and its verified neighbors are subsequently served as the starting points for the next expansion layer. For over-coordinated sites ($n_c > 4$), to resolve local clustering caused by interstitial noise, the algorithm systematically evaluates all possible four-neighbor combinations. The specific subset that yields the minimum CSM is selected as the true local structure, while the remaining unselected atoms are autonomously pruned as noise.

iii) Mathematical inference of missing vertices

For under-coordinated internal sites ($n_c < 4$), the spatial coordinates of the missing tetrahedral vertices are mathematically inferred based on the validated existing neighbors. For one missing vertex, the direction of the missing bond is strictly anti-parallel to the vector sum of the three known bond vectors. The theoretical coordinate $r_{theo}$ is defined as:

$$\boldsymbol{r}_{theo} = \boldsymbol{r}_0 - \bar{d} \frac{\sum_{j=1}^{3}(\boldsymbol{r}_j - \boldsymbol{r}_0)}{\left\|\sum_{j=1}^{3}(\boldsymbol{r}_j - \boldsymbol{r}_0)\right\|}$$

where $\bar{d}$ is the average bond length. For two vertices are missing, let the two known bond vectors be $\boldsymbol{u} = \boldsymbol{r}_1 - \boldsymbol{r}_0$ and $\boldsymbol{v} = \boldsymbol{r}_2 - \boldsymbol{r}_0$. The two missing vertices lie symmetrically on an intersection contour governed by the tetrahedral angle and the mean squared bond length $R^2 = (\|\boldsymbol{u}\|^2 + \|\boldsymbol{v}\|^2)/2$. This resolves to a linear system $\boldsymbol{Mp} = \boldsymbol{b}$, where $\boldsymbol{M} = [\boldsymbol{u}^T; \boldsymbol{v}^T]$ and $\boldsymbol{b} = [-R^2/3; -R^2/3]$**.** The exact theoretical coordinates are determined along the orthogonal normal vector $\boldsymbol{w} = \boldsymbol{u} \times \boldsymbol{v}$.(iiii) Refinement of the atom position

Theoretical coordinates ($\boldsymbol{r}_{theo}$) serve solely as initial position. To accurately determine the precise atomic positions, each inferred vertex is spatially refined to the local intensity maximum ($\boldsymbol{r}_{exp}$) within $V\,(\boldsymbol{r})$ via an 3D Gaussian fitting:

$$V\,(\boldsymbol{r}) = I_{bg} + I_{peak}\exp(-(\boldsymbol{r} - \boldsymbol{r}_c)^T \boldsymbol{A}(\boldsymbol{r} - \boldsymbol{r}_c))$$

where $\boldsymbol{A} = \boldsymbol{R}^T\boldsymbol{D}\boldsymbol{R}$ captures the anisotropic peak profile through rotation ($\boldsymbol{R}$) and scaling ($\boldsymbol{D}$) matrices. The refined coordinate $\boldsymbol{r}_{exp}$ must then pass strict validation criteria: its peak intensity ($I_{peak}$) must exceed the noise baseline, and it must maintain a minimum physical exclusion distance from all existing atoms. Subsequently, iterative cycles are employed to correct the positions of nearest-neighbor four-coordinated atoms. This local correction propagates to derive the complete atomic model. Consequently, the ALT algorithm can finally remove spurious atoms and restores the correct atomic configuration.

**Zinc-blende and Wurtzite (twin) polyhedral template matching**

From the traced atomic coordinates, we estimated the local atomic bond structures by using polyhedral template matching (PTM) in custom MATLAB code, in a similar manner to the procedure described elsewhere[59]. The basic premise of PTM is to define different crystal structures at each atomic coordinate by the similarity between the atomic coordinates formed by nearest neighbor shells and those from an ideal reference structure. The primary differences in our approach are: 1) We used a cost function which does not require an explicit mapping between pairs of experimental atomic coordinates and the reference structure. 2) We included up to ~5 neighboring shells of atoms up to radius of 13.2 Å to overcome experimental position uncertainty. 3) We matched the Wurtzite and Zincblende structures and their inverse structures, as both structures are not centrosymmetric.

To calculate the best-fit local structures, we first find all neighboring sites up to a maximum radius of 16.7 Å around each site and then subtract that site's position from its neighbors. We then loop through all pairs of neighboring atoms to find an initial guess for the rotation matrix which maps the experimental neighboring bonds to each reference structure using the Kabsch algorithm[60]. We then pair experimental sites with their nearest reference structure sites and further refine the rotation matrix and translations using another application of the Kabsch algorithm. Finally, we calculate an order parameter using the function

$$\mathrm{OP} = \frac{1}{N_{\mathrm{ref}}} \sum_{k=1}^{N_{\mathrm{ref}}} \max\left(1 - \frac{\min\left(\left\|r_{\mathrm{exp}} - r_{\mathrm{ref,k}}\right\|_2\right)}{r_{\ \mathrm{max}}}, 0\right)$$

Where $N_{\mathrm{ref}}$ is the number of reference sites, $r_{\mathrm{exp}}$ and $r_{\mathrm{ref}}$ are the experimental and reference coordinates respectively, and $r_{\mathrm{max}}$ is the maximum bond radius included which we set equal to 3.45 Å. Note that we find the minimum bond length from all experimental coordinates rather than defining an explicit pair mapping. We define the local crystal structure and orientation from the largest order parameter.

**Curvature calculations**

We have estimated the surface curvature of each atomic plane using the following procedure in custom MATLAB code. First, we extract all coordinates from a given atomic plane. Next, we interpolate a smooth surface[43]. We then apply Gaussian smoothing to the surface with a standard deviation equal to double the {002} plane spacing of the crystal structures to reduce the influence of experimental position errors. From this smoothed interpolated surface, we numerically compute each of the first derivatives $S_x$, $S_y$, and the second derivatives $S_{xx}$, $S_{yy}$, and $S_{xy}$ using center difference. We calculate the mean surface curvature $H$ using the expression

$$H = \frac{\left(1 + {S_x}^2\right)S_{yy} - 2S_xS_yS_{xy} + \left(1 + {S_y}^2\right)S_{xx}}{2\left(1 + {S_x}^2 + {S_y}^2\right)^{3/2}}$$

Finally, we estimate the surface curvature at each atomic coordinate from the nearest surface point.

**Pair distribution function**

Pair distribution function (PDF) of 3D atomic model of QDs was computed to characterize atomic spatial correlation[60,61], according to the following equation:

$$g(r) = \frac{\rho(r)}{\rho_0}$$

For each atom, pairwise distances were binned, and local densities $\rho(r)$ were acquired by dividing atom counts by the shell volume $V(r) = \frac{4}{3}\pi[(r + \Delta r)^3 - r^3]$. Boundary effect of atomic model was corrected by estimating the accessible volume fraction via alpha shape (MATLAB function) masking and spherical sampling. PDF was normalized by the global density $\rho_0 = N/V$, where N is the total number of atoms and V is the volume of QDs.

**Cation exchange**

To quantitatively determine whether an atom is a potential cation exchange site, we assessed the spatial distribution of all cations relative to their second-nearest neighbor cations with different atom types. Cation exchange is considered to occur at a site when the central atom is geometrically surrounded by neighboring cations of a different type. We began by screening all cations in the atom model to select those whose second-nearest neighbors included cations of a different chemical species, indicating potential sites for cation exchange. Then, we developed a directional coverage metric based on the angular distribution of neighbor vectors to evaluate the spatial geometry between central atom and its neighbors. This is formulated as a constrained optimization problem:

$$\min_{||v||=1} \max_{i}(P_i \cdot v)$$

Where $P_i$ is the unit vector pointing from the center to each neighbor. We simplify this problem by identifying the direction $v$ that maximizes the minimum angular separation to all neighbor vectors. The quantity $\theta_{max} = cos^{-1}(max_i(P_i \cdot v))$ represents the largest angular aperture in which no neighboring atom is present. If $\theta_{max}$ is smaller than 90°, the central cation is considered to be fully surrounded by cations of a different type, demonstrating that this site has undergone cation exchange.

**Orientation calculation and isosurface/shape size estimation**

For the orientation of QDs, a normal vector at each atomic site was calculated by evaluating the gradient of the local density matrix. This established method providing a robust way to characterize surface facets at the atomic scale has been published elsewhere. By comparing the angular relationships between the calculated normal vectors and known crystallographic directions, we were able to accurately determine the facet orientations of the InAs core and ZnSe shell in InAs/ZnSe QDs[50].

**Low-magnification electron tomography**

Electron tomography of InAs/ZnSe QDs at low magnification was carried out on aberration corrected Thermo Fisher Scientific Spectra Ultra S/TEM microscope operated at 300 KV. ADF projections were acquired over a tilt range of –69.90° to 69.00°, using a convergence angle of 17.9 mrad and collection angle of 35 to 200 mrad. At each tilt angle, three consecutive projections were recorded at a magnification of 1.80 Mx (pixel size: 1.086 Å) with a dwell time of 3 μs, resulting in a total electron dose of $1.76 \times 10^4$ $e^-/Å^2$. To accurately reconstruct the 3D volume of InAs/ZnSe QDs, the tilt series underwent a preprocessing pipeline established for AET above. Specifically, the raw projections were subjected to sequential drift correction, image denoising, and background subtraction to maximize the signal-to-noise ratio. After precise alignment to a common tilt axis, the 3D tomogram was reconstructed utilizing the RESIRE algorithm, as detailed in the previous section.

**DFT calculations**

All calculations were performed using periodic Density Functional Theory (DFT) implemented in the Quickstep module of the CP2K software[62,63], utilizing the Gaussian and Plane Wave (GPW) method. Structural optimization for the quantum dot systems was carried out using the DZVP-MOLOPT-SR-GTH molecularly optimized basis set along with Goedecker-Teter-Hutter (GTH) pseudopotentials[64,65]. To overcome the well-known tendency of standard Kohn-Sham density functional theory (DFT) with local or semi-local functionals, such as the local density approximation (LDA)[66] or generalized gradient approximation (GGA)[67], to underestimate bandgaps due to self-interaction errors and the absence of derivative discontinuity, we employed the PBE0 hybrid functional[68]. By incorporating a fraction of exact Hartree-Fock exchange, PBE0 provides a more accurate description of the electronic structures of semiconductor nanostructures. The energy convergence criterion for the self-consistent field (SCF) calculations was set to $5\times10^{-6}$ Hartree, with a plane wave energy cutoff of 400 Ry. To minimize interactions between adjacent periodic images, a vacuum slab of 10 Å was introduced[69]. Building on the ground state calculations, we addressed electron-hole interactions through Time-Dependent DFT (TDDFT), a predictive framework for determining the optical gaps of colloidal quantum dots (QDs). Given the significant scale of the model system (over 1,200 atoms), which renders conventional TDDFT computationally prohibitive due to its scaling, we utilized the simplified TDDFT (sTDDFT) approach[70]. By employing a multipole approximation for the transition density, sTDDFT captures the essential physics of PBE0-based electronic transitions with high fidelity to standard TDDFT at a fraction of the cost. This methodology allowed us to resolve the 50 lowest-energy excitonic states and accurately simulate the low-energy absorption lineshapes of these large-scale systems. Input files and data processing were handled using Multiwfn 3.8 (dev)[71,72].

Due to the high computational cost, simulating a full core/shell QD exceeding 10,000 atoms is prohibitive using hybrid density functional theory (DFT). However, the type-I band alignment[73] and sharp interface of InAs/ZnSe naturally localize charge carriers within the InAs core, which remains the primary region of interest for optical properties. By extracting the InAs core directly from the AET-resolved structure and effectively passivating surface dangling bonds, we retain realistic internal features and heterogeneous shell-induced strain while reducing the system to a tractable size. We refer to this as the experimental QD (exp. QD). To isolate the effects of shell-induced strain, we generated a relaxed model via full geometry optimization of the exp. QD. While the exp. QD inherits the heterogeneous strain and geometric constraints of the parent ZnSe shell, the relaxed model represents an energy-minimized state where these constraints are absent. Finally, a nearly spherical QD truncated from a zinc-blende InAs bulk crystal was used as an ideal reference (ref. QD). By neglecting shape anisotropy, strain variations, and twin boundaries in this reference model, we demonstrate the necessity of AET-resolved realistic structures for accurately reproducing experimentally measured optical properties. In all those QD systems, the surface dangling bonds are fully passivated using the pseudo-hydrogens with fractional charges of 1.25 e and 0.75 e for passivating surface cation and anion atoms, respectively.

To compute the density of transitions (DOT), a Gaussian broadening function was applied to the calculated excitation energies ($E_{ex}$), and a summation over the Gaussian distribution was performed using the function:

$$DOT = \sum_{ex} \frac{1}{\sqrt{2\pi\sigma^2}} \exp\left(-\frac{(E - E_{ex})^2}{2\sigma^2}\right)$$

For those calculations, a broadening parameter of $\sigma$ = 50 meV was chosen[74].

We used the Multiwfn software package to calculate the dipole moment of electrons in the QD system. The calculation formula was as follows,

$$p_A = -\int \omega_A(r)\,\rho(r)dr$$

Where $\rho(r)$ is the electronic density, and $\omega_A(r)$ is the Hirshfeld atomic weight function of atom $A$ represents the set of all atoms in the system.

We examined the inverse participation ratios (IPRs) of a given electronic or excited state. The IPR was calculated on the basis of a basin analysis of real-space densities of a given electronic state or attach (or detach) density for a given excitonic wave functional using the Multiwfn software, and it estimated the number of attractors contributing significantly to each involved density, which was defined as:

$$IPR^{\emptyset} = \frac{1}{N\sum_A \left(q_A^{\emptyset}\right)^2}$$

Where $\emptyset$ denoted the HOMO or LUMO orbital, $q_A^{\emptyset}$ was the contribution of attractor A, and N was the normalization factor[74,75].

**Setups for DFTB geometric relaxation**

We employed the Density Functional Tight Binding (DFTB) framework[45] to perform geometric relaxation on InAs/ZnSe core/shell quantum dots (QDs) containing over 12,000 atoms. This scale renders standard

DFT calculations computationally prohibitive. To ensure high efficiency and accuracy, we utilized s and p atomic shells for all elements, with the addition of the d shell for the heavier Zn atoms. The Slater-Koster parameter file for all elements are taken from Margraf et., al.[76]. The DFTB framework was first validated against standard DFT calculations using the CP2K software suite[62,63], employing the PBE functional and a DZVP-MOLOPT-SR-GTH basis set[64,65]. This validation yielded a relative bond-length error of less than 2%, confirming the structural accuracy of our approach. The realistic QD model, totaling 12,472 atoms, was derived from an AET-resolved atomic structure by removing the outermost disordered layer and passivating dangling surface bonds with pseudo-hydrogen atoms to ensure self-consistency. Simulations were conducted on a supercomputing facility using 768 cores across four nodes (each equipped with 192 cores and 3 TB of RAM). Given the characteristically flat potential energy surface of such large-scale systems, convergence criteria for energy and force were set to $10^{-3}\ a.u.$ and $2 \times 10^{-2}\ a.u.$, respectively. The final structural optimization reached completion in approximately 100 iterations.